\documentclass[final,5p,times,twocolumn]{elsarticle}

\usepackage{amssymb}

\usepackage{lineno}

\usepackage{graphicx}
\usepackage{morefloats}
\usepackage{subfigure}
\usepackage{amsmath}
\usepackage{amssymb}
\usepackage{amstext}
\usepackage{epstopdf}

\biboptions{comma,square,sort}

\journal{Nuclear Instruments and Methods A}

\begin{document}

\begin{frontmatter}



\title{The Magnetic Distortion Calibration System of the LHCb RICH1 Detector}


\author[su]{A.~Borgia}
\author[ic]{W.~Cameron}
\author[ox]{A.~Contu\fnref{au4}}
\author[ce]{C.~D'Ambrosio}
\author[ce]{C.~Frei}
\author[ox]{N.~Harnew}
\author[ox]{M.~John}
\author[su]{G.~Lefeuvre\fnref{au3}}
\author[su]{R.~Mountain\corref{cr1}}
\author[su]{S.~Stone}
\author[ic]{D.~Websdale}
\author[ox]{F.~Xing}

\address[ce]{European Organization for Nuclear Research (CERN), CH-1211 Geneva 23, Switzerland}
\address[ic]{Imperial College London, Physics Department, London SW7 2AZ, UK}
\address[su]{Syracuse University, Department of Physics, Syracuse NY 13244, USA}
\address[ox]{University of Oxford, Denys Wilkinson Building, Oxford OX1 3RH, UK}

\cortext[cr1]{Corresponding author. Email address: {\tt raym@physics.syr.edu}.}
\fntext[au3]{Now at: University of Sussex, Brighton BN1 9RH, UK}
\fntext[au4]{Now at: INFN Cagliari, 09042 Monserrato, Italy}

\begin{abstract}

The LHCb RICH1 detector uses hybrid photon detectors (HPDs) as its optical sensors.  A calibration system has been constructed to provide corrections for distortions that are primarily due to external magnetic fields.  We describe here the system design, construction, operation and performance.

\end{abstract}


\end{frontmatter}


\section{Introduction}

The LHCb experiment~\cite{LHCbDetector} is dedicated to precision measurements of CP violation and rare decays of B hadrons at the Large Hadron Collider at CERN. Key measurements of the LHCb physics program are outlined in \cite{LHCb-PUB-2009-029}.  Excellent discrimination among different particle species is vital to study decays with similar topologies and to suppress combinatorial backgrounds.

LHCb has two RICH detectors to provide hadron identification in the momentum range 2--100 GeV/$c$.
Both RICH detectors use arrays of hybrid photon detectors (HPDs) to detect the Cherenkov photons created in their radiators~\cite{LHCbDetector,RICH-TDR,LHCb-TDR-R}.  The photon detectors need to provide accurate photon position measurements.  This is a critical component of the Cherenkov angle resolution, which determines the separation power of different particle species, particularly at high momentum.  Indeed, at all momenta, good photon position resolution is required by the likelihood function that is typically used in particle identification algorithms.

The HPD used in the LHCb RICH detectors~\cite{Gys01,Moritz04} is an optoelectronic imaging vacuum tube that has a quartz window of spherical section, an S20 multi-alkali photocathode, and a silicon pixel sensor.
Its internal electrodes are operated at 16--20 kV, setting up an electric field cage in which the photoelectrons are cross-focused, and follow long ($\sim$125 mm) drift trajectories from the photocathode to the pixel sensor.
The pixel sensor is bump-bonded to a custom binary readout chip.
There are 32$\times$32 effective pixels in the sensor, each of area 0.5$\times$0.5 mm$^2$.
However, each effective ``LHCb'' pixel actually consists of
eight physical ``ALICE'' pixels of area 0.5$\times$0.0625 mm$^2$.  The eight ALICE pixels are grouped together in the readout to form a single LHCb pixel (termed ``pixel'' herein).
The nominal electrostatic demagnification factor of five gives a measurement granularity of 2.5$\times$2.5 mm$^2$ at the exterior of the quartz window.

HPDs are well known to be sensitive to external magnetic fields~\cite{AR2004,AR2005,AR2006}.  In particular, they are efficient only below 1.5~mT axial fields, and less so if the field is off-axis.  The magnetic Lorentz force deflects photoelectrons from their nominal trajectories causing pinwheel and shift distortions of the entire image and hence a subsequent loss of accuracy in the detected photon position.  Depending on the angle of the magnetic field, this distortion can be quite severe (up to several pixels), completely compromising the photon position resolution.

\begin{figure}[htb]
\centering
\includegraphics[width=6cm,bb=0 0 447 535]{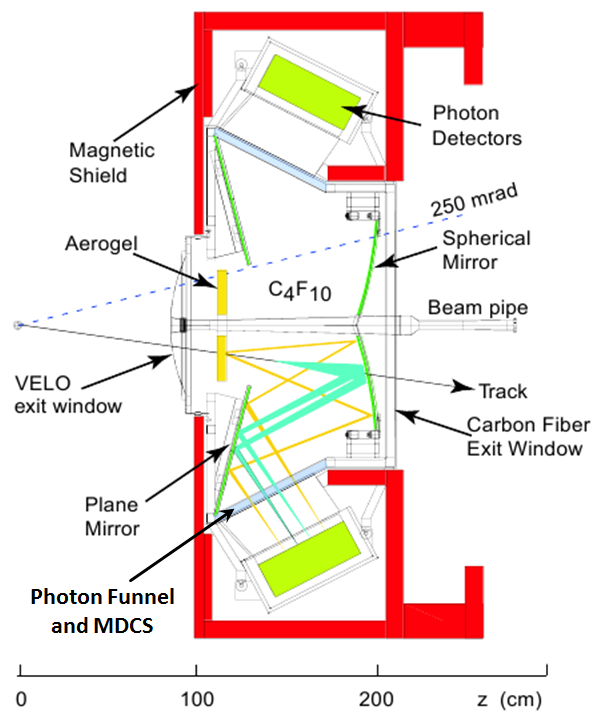}
\caption{Schematic view of the RICH1 detector.  The MDCS system sits in the photon funnel, which is located just outside of the gas enclosure.  There are two MDCS systems, one for each of the upper and lower HPD arrays. Each HPD array has 7 columns of 14 HPDs each.}
\label{fig:R1_layout}
\end{figure}

Both LHCb RICH detectors are affected by the fringe magnetic field of the spectrometer magnet, but it is most pronounced in the RICH1 detector which is located immediately upstream of the magnet.  The field is reduced by an iron shield surrounding RICH1 (seen in Fig.{~\ref{fig:R1_layout}}), and further by cylindrical Ni-Fe alloy shields around each HPD.  Even so, initial estimates indicated up to 3.0~mT fields at the HPDs.  With possible inhomogeneities and saturation effects, the resulting non-uniform fields could produce serious distortions which would be impossible to predict in an a priori manner.  The calibration system described in this paper is dedicated to mapping and correcting these magnetic distortions in RICH1~\cite{Xing2011zz}.  A different system is employed for this purpose in RICH2~\cite{Cardinale2011eu}.  In addition to magnetic distortions, there are electrostatic distortions possible from variations in the applied HPD high voltages, and optical distortions from the quartz window.  Distortions from all sources are combined, and this calibration system will correct for these effects as well.

In the next section, the overall design of the magnetic distortion calibration system (termed MDCS herein) is provided, followed by a description of the relevant details of the system, its testing and installation.  Subsequently, the data-taking procedure and analysis method are developed, then the main performance metrics and results are presented.  Concluding remarks follow.

\section{System Design}

The basic requirement on the MDCS system is to remove the potentially severe distortions from the array of HPDs.  The corrected photon position should contribute an error small in comparison with the intrinsic spatial resolution of the HPD.

The strategy for the MDCS is implemented in two steps.  First, a light spot is projected at a precisely-known position on the quartz window of a given HPD in the array.
The spot has a small size in comparison to the pixel size, and is oriented front-on to minimize refractive effects at the window and shadowing by the magnetic shields.  The spot is moved to scan the HPD array in two dimensions, with small enough step sizes to map out the distorted image of the pixels.  The data taken from this scan form a ``direct'' mapping of the distortion at a given magnetic field.  Second, an ``inverse'' mapping is created, assigning a given distorted pixel hit to its real position on the photocathode.  This may be done by a parameterized functional fit to the data if the distortion is sufficiently smooth to do so with a small enough reconstruction error, or alternately by a look-up table if the distortion is too severe or non-uniform.  With proper survey information, this strategy can provide an absolute positional calibration of each pixel in the HPD array.

Implementing this strategy for the MDCS system required consideration of the tight physical constraints on the system in terms of overall area (1283$\times$540 mm$^2$), form factor ($<$40 mm high), and maintenance of a clear photon aperture for the HPD array.  Additional considerations were made due to operation in a magnetic and radiation environment.

\begin{figure}[thb]
\centering
\includegraphics[height=8cm,bb=0 0 526 650]{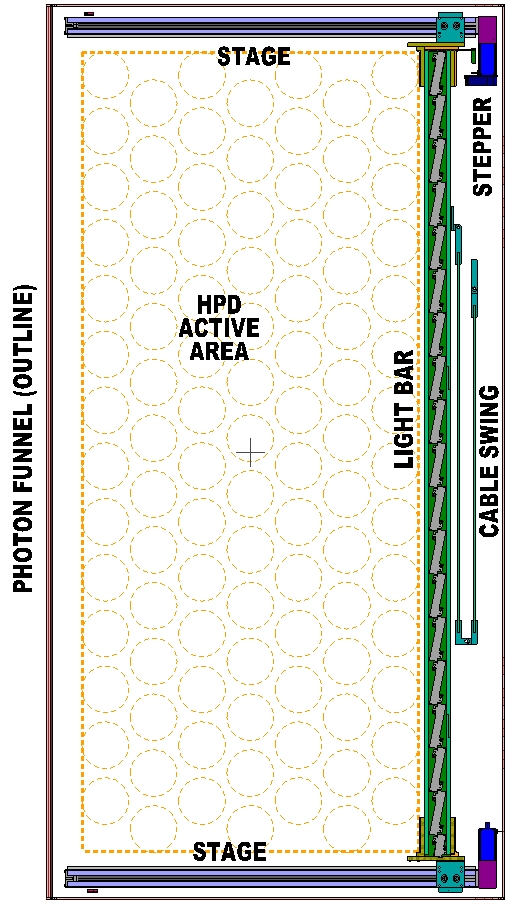}
\caption{MDCS overall system drawing.  In this view, HPDs are located above the system, facing down (into page).  Their positions are denoted by dotted circles.  The light bar movement is horizontal, and it is shown in its fully-retracted (parked) position.}
\label{fig:system_schema}
\end{figure}

The basic system design is shown schematically in Fig.{~\ref{fig:system_schema}}.
Instead of a single light spot moving in two dimensions, spatial constraints on the system size forced a solution having a long bar with many individual light sources (LEDs), termed a ``light bar.''  Each LED in the light bar can be individually powered, and each is collimated in order to project a small light spot onto the HPD focal plane.  The light bar is mounted as a gantry between two linear-motion control stages.  The movement of a given light spot in the direction of the stages is very fine and nearly continuous, while the positioning of the light spot in the direction along the light bar is discrete, effected by powering adjacent LEDs in turn.  This choice is dictated by the spatial constraints on the system.

\section{System Description}

Two identical and independent MDCS systems were designed and constructed for RICH1, one for each of the upper and lower HPD enclosures.  In this section, the description pertains to only one of the two systems.

\subsection{Light Bar}

The light bar consists of two long carbon fiber side rails on which are mounted a series of 19 LED PCB
units, custom designed for this application\footnote{Design and fabrication of these LED PCB units (and controller) was made in conjunction with SenSyr LLC, 111 College Place, Syracuse NY 13244, {\tt www.sensyr.com}.}.

\begin{figure}[htb]
\centering
\includegraphics[height=5cm,bb=0 0 1064 700]{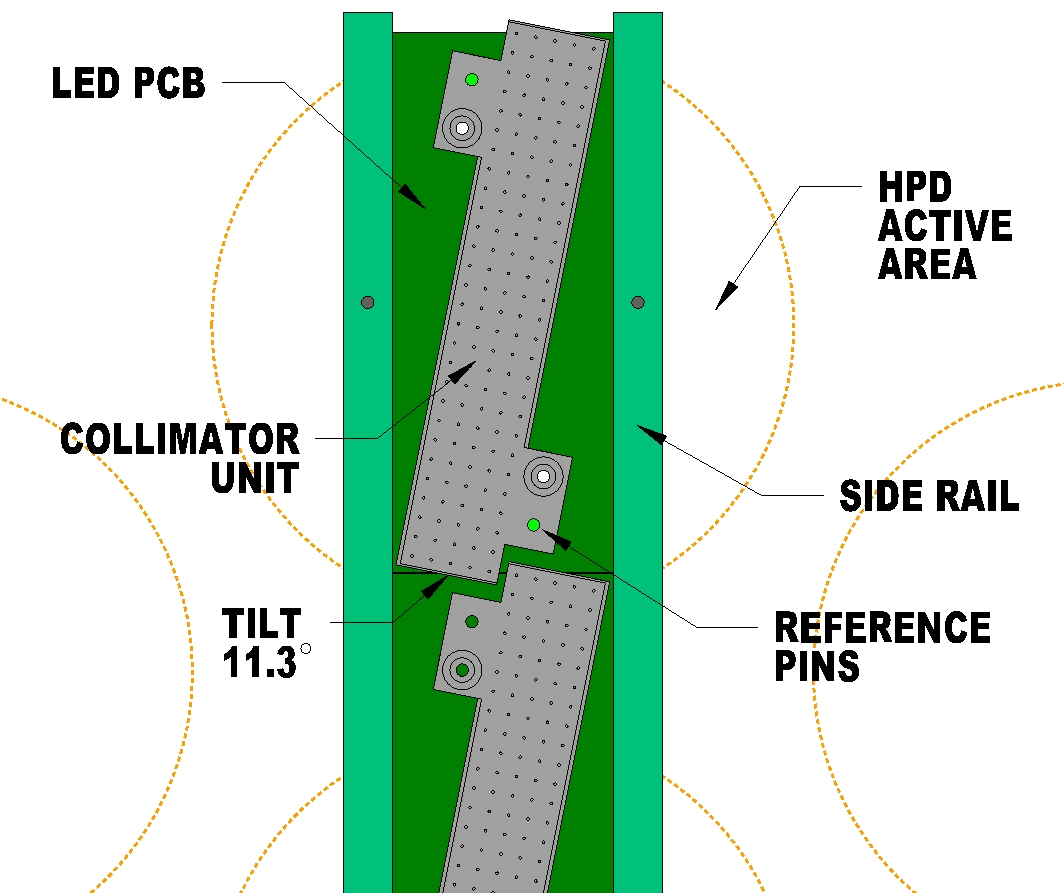}
\caption{A close-up of the light bar design, showing a full LED PCB unit, as viewed from the HPD array.}
\label{fig:led_pcb}
\end{figure}

Each LED unit consists of a six-layer PCB, on which are mounted four LED arrays, an on-board addressable microcontroller, a driver circuit to control each LED, and a collimator unit.  This unit is shown in Fig.{~\ref{fig:led_pcb}}.

The LED arrays\footnote{Lite-On LTP-757G LED arrays.} each consist of a $5\times7$ matrix of green LEDs on 2.54~mm centers.  The LEDs have dominant wavelength 569 nm and half-width 30 nm.  The four LED arrays are tilted as a unit at an angle of $\sin^{-1}(0.5/2.54)=11.4^\circ$, in order to provide a smaller effective separation of light sources.  This tilt gives a 0.5 mm granularity of LED centers, as projected in the long dimension of the light bar.
During construction, the LED arrays were precisely aligned to reference holes in the PCB, epoxied in place, then soldered.
The PCB was populated afterwards with the remaining components.

The microcontroller\footnote{Microchip microcontroller PIC 18F6622, with high current source/sink.} on each PCB is addressed individually and downloaded with a pattern code via an SPI bus\footnote{Serial Peripheral Interface (SPI) bus is a de facto standard for device communication and control in master/slave mode, via synchronous serial data link.}
which runs along the light bar.  To power an LED, a microcontroller output line goes high, sourcing the current for one or more selected rows of five LEDs.  Concurrently, the microcontroller fires one or more column power MOSFETs\footnote{International Rectifier IRLML2803 HEXFET Power MOSFET.} to sink the current.  The pattern code selects which row(s) and column(s) are activated.  All necessary LED array patterns may be activated with this logic. Each LED PCB may be programmed with a different pattern or all with the same pattern.

\begin{figure}[tbh]
\centering
\includegraphics[height=4cm,bb=0 0 779 700]{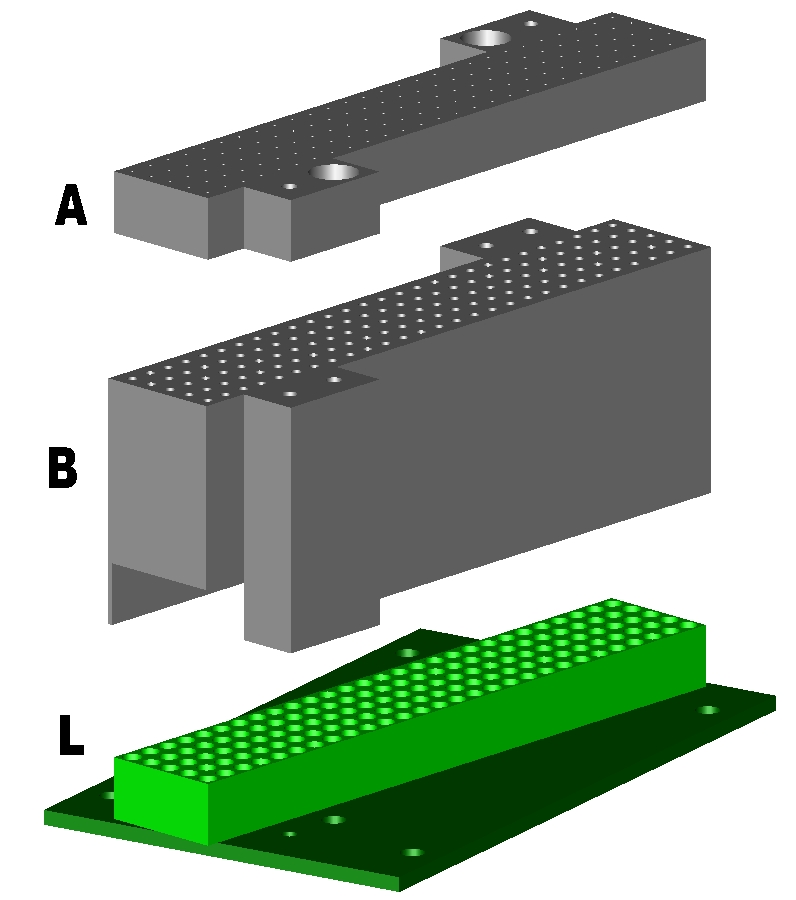}
\caption{An exploded view of a single LED PCB unit.  Light is emitted from LED array L into the collimator unit, encountering first part B having short small-diameter holes followed by longer larger-diameter holes, and then part A having the same short small-diameter holes.  Light emerges upwards toward the HPD array (not shown).}
\label{fig:led_pcb_exp}
\end{figure}

The collimator unit is an array of individual collimators, one for each LED in the LED array, shown in Fig.{~\ref{fig:led_pcb_exp}}. Each collimator is centered on an LED, and consists of a length of open tube machined in a block of Delrin$\circledR$ acetyl homopolymer.  The collimator geometry has two 6.35 mm lengths of 0.343 mm diameter separated by one 12.7 mm length of 1.0 mm diameter. Extensive design studies of the collimator geometry were carried out, with the projected spot size as the figure of merit.  The spot size was designed to be 6.5 mm FWHM at the effective HPD focal plane, 85 mm from the collimator.

A practical consideration was the ability to reliably and precisely machine an array of 2590 identical collimator holes in a light bar.  This influenced the decision to use Delrin$\circledR$, which is suited for CNC machining in its ability to be cut without binding the tools, and which has low moisture retention hence good dimensional stability over time.  The collimator unit was constructed as a stack-up of two pieces, due to the limitations in the thickness of material though which a small-bore hole can be drilled reliably.  These two pieces were visually inspected after manufacture, and re-worked by hand as necessary to assure each small hole was free of any residual obstructions.

Mechanically, the collimator unit is the precision reference point for fabricating the light bar.  The collimator unit was assembled on the populated LED PCB using the same reference holes as the LED arrays.  A precision jig was used to align all 19 completed LED PCBs\footnote{The last LED PCB in the light bar is half-sized, containing only two LED arrays.  All other 18 LED PCBs are full-sized.} in the light bar by locating the exit faces flat on a reference plane and transversely by using these same PCB reference holes transferred to the face of the collimator units.  The result was a $\sim$1200 mm long, planar, two-dimensional array of 2590 collimator holes precisely located on 0.5-mm projected centers, across LED and PCB boundaries, with reference marks for surveying into RICH1.

\subsection{Motion Control}

Motion control was effected by two linear stepper motor stages\footnote{Parker/Daedel model LP28 linear-motion control stages.}, with a long enough travel to cover the entire active area of the HPDs.
The stepper motors allow for precision movement of the light bar.  An optical position encoder gives feedback position information for additional accuracy. Both stages are wired to a single motor controller, synchronizing the movement of both stages.  Each stage was modified with a 90-degree helical gear box to fit the tight spatial constraints.  The gearbox also allows for manual turning from outside of RICH1, should the stages fail.

Each stage is fitted with two limit switches which indicate the home and end-of-travel positions.  During movement, a swing mechanism keeps the light bar cabling from binding.  Mounting pins allow the removal and replacement of the stages with precision.  Similarly, the light bar is mounted with locating pins from one end for precision reference.

\begin{figure}[thb]
\centering
\includegraphics[height=6cm,bb=100 0 680 650]{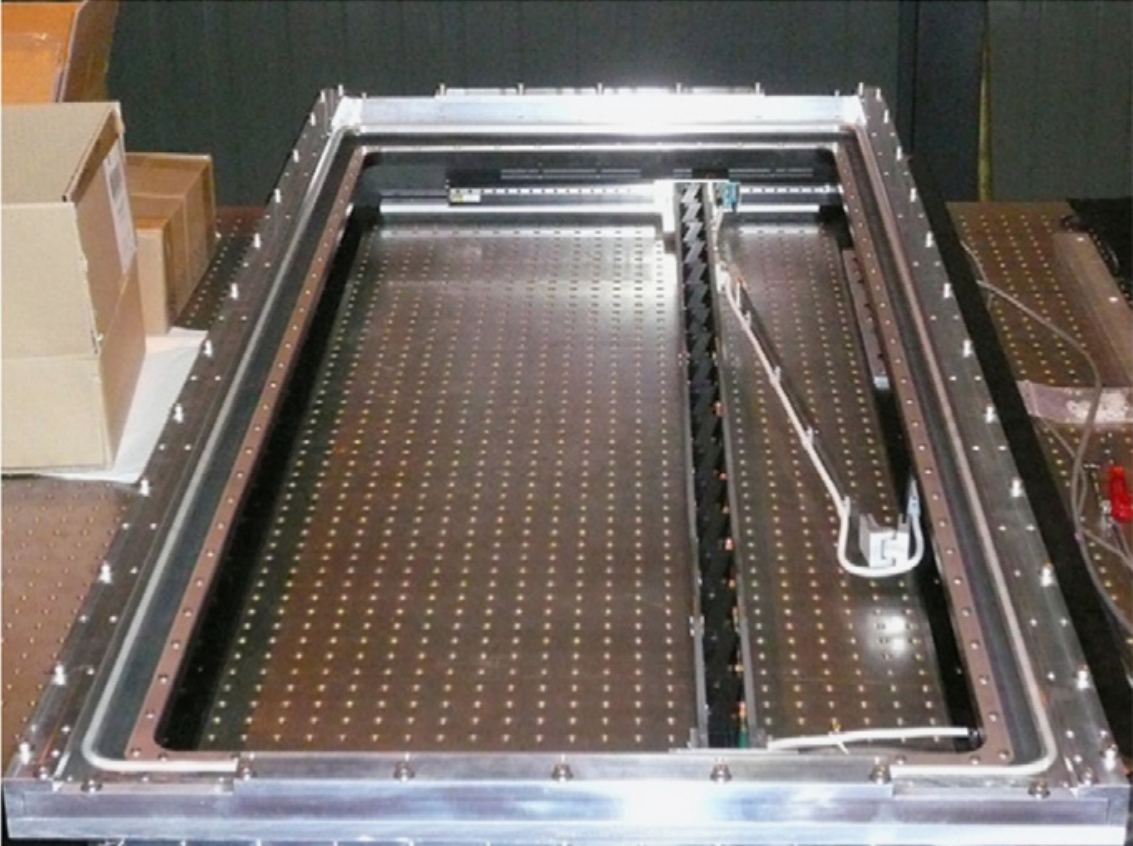}
\caption{A photograph of one MDCS system, installed in a photon funnel prior to installation in RICH1.  The light bar is seen extended at $\sim$1/3 of its full travel.  During RICH operation, it is folded away into its parked position.}
\label{fig:system_photo}
\end{figure}

\subsection{Master Control Board}

The light bar and motion stages are both controlled by the master control board, a custom PCB which has a master microcontroller, internet interface, LVDS drivers, power drivers, and AC relay.  The master microcontroller\footnote{Microchip microcontroller PIC 24FJ128GA006. This is the master to the slave microcontrollers which are on the LED PCBs.} was programmed with its own small operating system to respond to specific commands arriving from the internet port\footnote{Lantronix XP1002001-03R XPort SE embedded ethernet server.}.  This allows all commands to be issued via TCP/IP.  Depending on the command, the master control board will either pass the command to the LED PCBs, pass it to the stepper motor controller, or act on the command itself.  Hence by synchronizing the LED pattern and movement of the light bar, a cartesian grid pattern (or other required pattern) may be projected on the entire HPD array. 

The light bar commands are transmitted to the LED PCBs over a long cable using LVDS logic to reduce noise pickup, and received by a transition board at the head of the light bar.  Here the signals are level-adapted to the TTL SPI bus which communicates to the slave microcontroller on each LED PCB individually or collectively.

The master control board can also turn off the DC power to the light bar and AC power to the motor controller.  This function is quite important, since even when idle but powered, these devices may be a source of noise for the HPDs during operation.

\begin{figure}[thb]
 \centering
 \includegraphics[width=6cm,bb= 0 0 567 539]{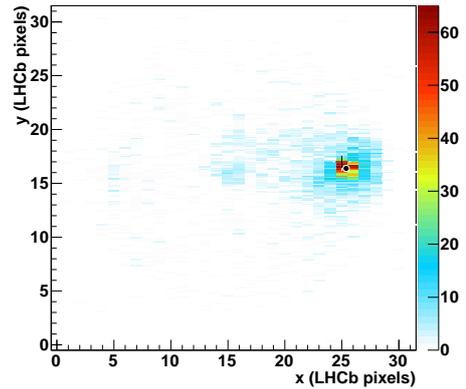}
 \caption{A pixel hit map of a single LED light spot (with ALICE pixels in the y direction). The black dot marks the center of a 2D Gaussian fit, and the cross marks the position of the arithmetic mean, for comparison.  The smattering of other hits are from background processes, as discussed in the text.}
 \label{fig:example_hitmap_good}
\end{figure}

\section{Testing and Installation}

Both MDCS systems were tested extensively for mechanical and optical performance.

The motion control stages have a position resolution of 50 $\mu$m bidirectionally.  However, a repeatability of $\sim$10 $\mu$m is obtained when traveling unidirectionally (i.e., when the movement includes no backlash in the lead screw).  So in operation, any such effect is reduced by always approaching a new position for the light bar from the same direction.

The light output of each spot was set via a current-limiting resistor to have a 1 MHz count rate as measured by the HPD at the center of the spot, which is equivalent to the highest rate expected in actual operation.

Both MDCS systems were mounted in their respective photon funnels before installation in RICH1, as shown in Fig.{~\ref{fig:system_photo}}.  The positions of the light bar at several points along its travel were surveyed with respect to the photon funnel.  After installation, the photon funnel was surveyed with respect to the RICH1, and thence defined with respect to the LHCb spectrometer.

\section{Operation in RICH1}

Special data-taking runs (called MDCS scans) are taken periodically for monitoring purposes, or in the event a new calibration is needed (e.g. HPD replacement).  Data are taken with both polarities of the spectrometer magnet (at 1.5 T nominal) as well as with the magnet off.

The light bars are controlled interactively via TCP/IP using a PVSS project\footnote{PVSS process control software, from ETM Professional Control GmbH, Eisenstadt, Austria.}, integrated into the RICH1 global control system.
The LEDs are synchronized with light bar movement to create a grid pattern of light spots over the entire HPD array.
A scan is divided into ``steps,'' uniquely characterized by a fixed bar position and a given configuration of LEDs. For simplicity, all LED PCB units are configured identically, with only one LED per unit switched on.

During a scan, the light bar moves 42 times and illuminates each HPD column, in turn, for $\sim$5 sec.
Typically, the sparse LED pattern results in an HPD being illuminated by a single LED at each step.
In this way, the pixel hits can be directly associated with a specific LED position.
For a given light spot, several pixels are hit.  The hit pixels are grouped together
into a cluster, called herein a ``peak'' (as illustrated in Fig.{~\ref{fig:example_hitmap_good}}).

\section{Calibration Data Analysis}

The first tasks in analyzing calibration scan data are to identify the peaks (clusters), and to clean the images in order to distinguish peaks corresponding to LED light spots from those due to unwanted background processes.

\begin{figure}[t]
 \centering
 \includegraphics[width=6cm,bb=50 0 500 384]{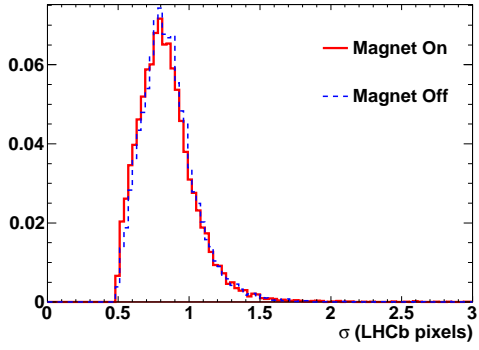}
 \caption{The radial width $\sigma$ of cleaned peaks, for magnet on data (solid red) and magnet off data (dashed blue).}
 \label{fig:sig_offon}
\end{figure}

\begin{figure}[t]
\begin{center}
 \subfigure{\includegraphics[height=6cm,bb=50 0 500 550]{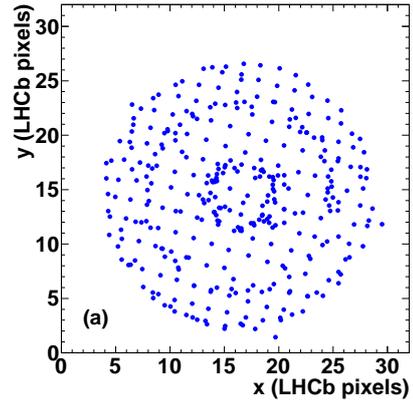}
 \label{fig:all_peaks}
 }
 \subfigure{\includegraphics[height=6cm,bb=50 0 500 550]{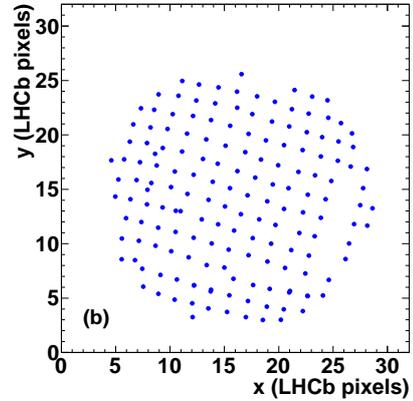}
 \label{fig:final_peaks}
 }
 \caption{Fitted peak patterns in an HPD image.
 (a) Typical image, before cleaning, with both signal and background peaks visible.
 (b) Same image, after cleaning, and before distortion correction.  Only signal peaks remain.}
 \label{fig:cleaned_peaks}
\end{center}
\end{figure}

\subsection{Peak Finding}

The peak-finding algorithm forms peak candidates from the photoelectron hits occurring during the 6000 triggers that are recorded at each step.
Pixels with more than 100 hits/step and with one or more of the eight neighboring pixels having no hits, are flagged as noisy, and removed from the analysis.

Peak candidates are formed in a square window of 7$\times$7 LHCb pixels
where the central pixel contains at least 1\%\ of the total hits in the
HPD (for that particular step) and the neighboring pixels have fewer entries.
Overlap between these matrices from different peaks is not allowed.
Selected peak candidates are then fitted using a two-dimensional Gaussian function.

The peak width distributions
of the width of the peaks
for magnet off and on data are compared in Fig.{~\ref{fig:sig_offon}}, and show that the peak shapes are not significantly distorted by the magnet.
The average peak radius is 0.84 pixels which corresponds to a light spot of 5.5 mm FWHM at the photocathode, to be compared to the expected design value of 6.5 mm FWHM.

\begin{figure}[b]
 \centering
 \includegraphics[width=6cm,bb=100 0 500 400]{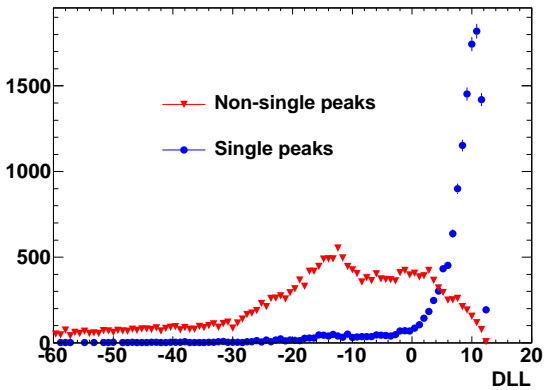}
 \caption{DLL distribution of the peak fitting, showing single peaks (blue circles) and non-single peaks (red triangles).}
 \label{fig:MDMSDLLdistrib}
\end{figure}

\subsection{Sources of Background}

Three independent sources of background contribute fake peaks in the scan data:
ion feedback~\cite{Eisenhardt2010391},
total internal refraction within the photocathode window,
or noisy pixels in the silicon chip.

Ion feedback occurs when a photoelectron ionizes residual gas atoms in the HPD tube. The ion drifts to the photocathode producing secondary electrons which are focused on the sensors with a typical delay of 200--300 ns. Usually the fake peaks due to ion feedback are located near the chip center.  This background occurs only in certain tubes in the HPD array.

Backgrounds from total internal reflection occurs whenever an LED illuminates the periphery of the quartz window.
Light entering the tube may be reflected back into the quartz window by the flat surface of the HPD dynode.
It is then incident on the external surface of the window at a sufficiently large angle for total internal refraction to occur. The curvature of the window dictates the radii at which the reflected-then-refracted light forms ancillary peaks. This background occurs at a similar radius, regardless of azimuthal angle, and produces the circular features visible in Fig.{~\ref{fig:all_peaks}}.
This background occurs in all HPDs.

HPDs with noisy pixels or those affected by ion feedback can be easily identified since they always have hits, even when not illuminated.  These tubes usually have a total number of reconstructed peaks that is well above the average of other tubes, integrated over the entire scan.

\begin{figure}[bth]
 \begin{center}
 \subfigure{\includegraphics[width=6cm,bb=50 0 550 500]{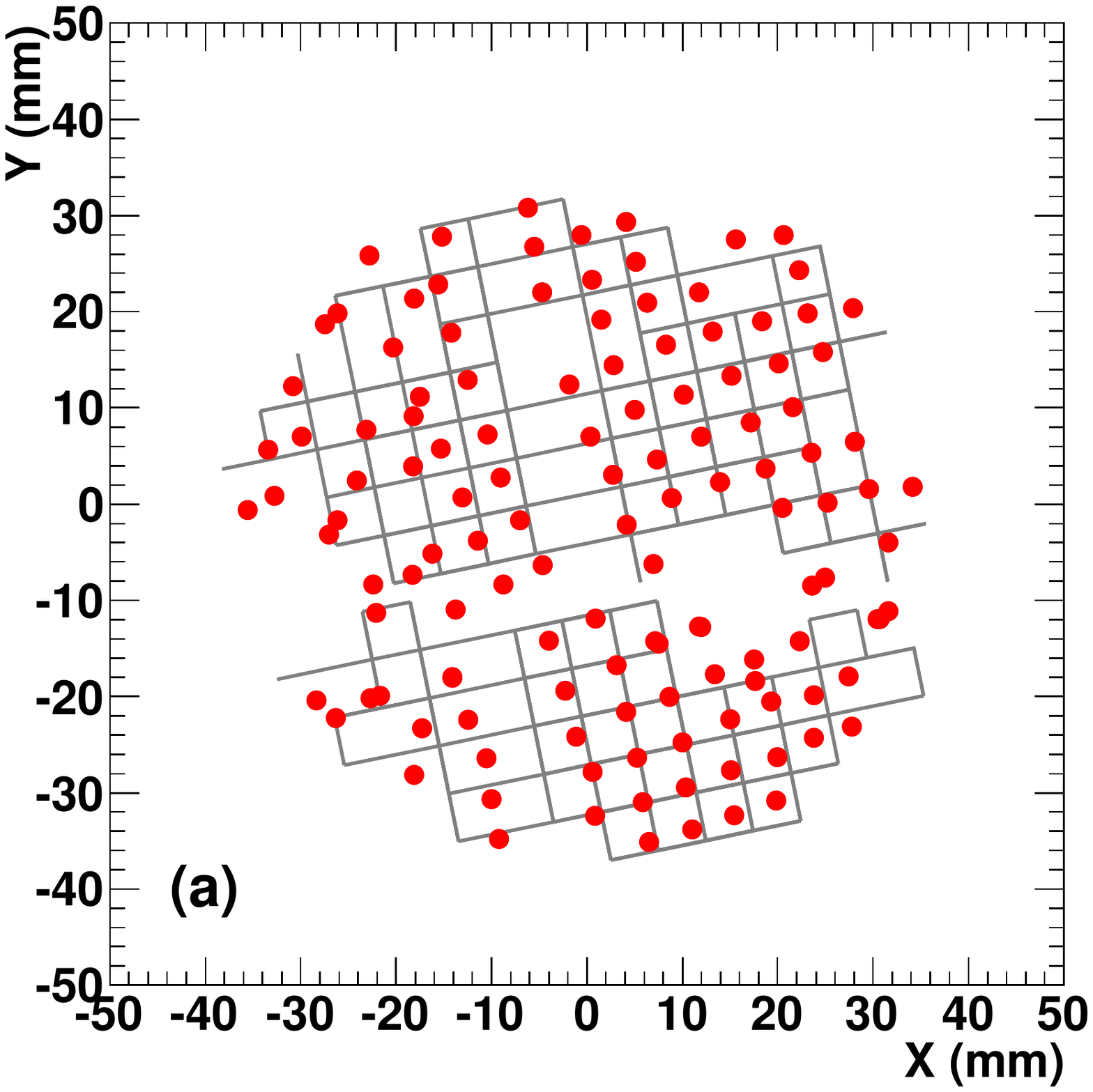}
 \label{fig:parfit_uncorr_2grids}}
 \subfigure{\includegraphics[width=6cm,bb=50 0 550 500]{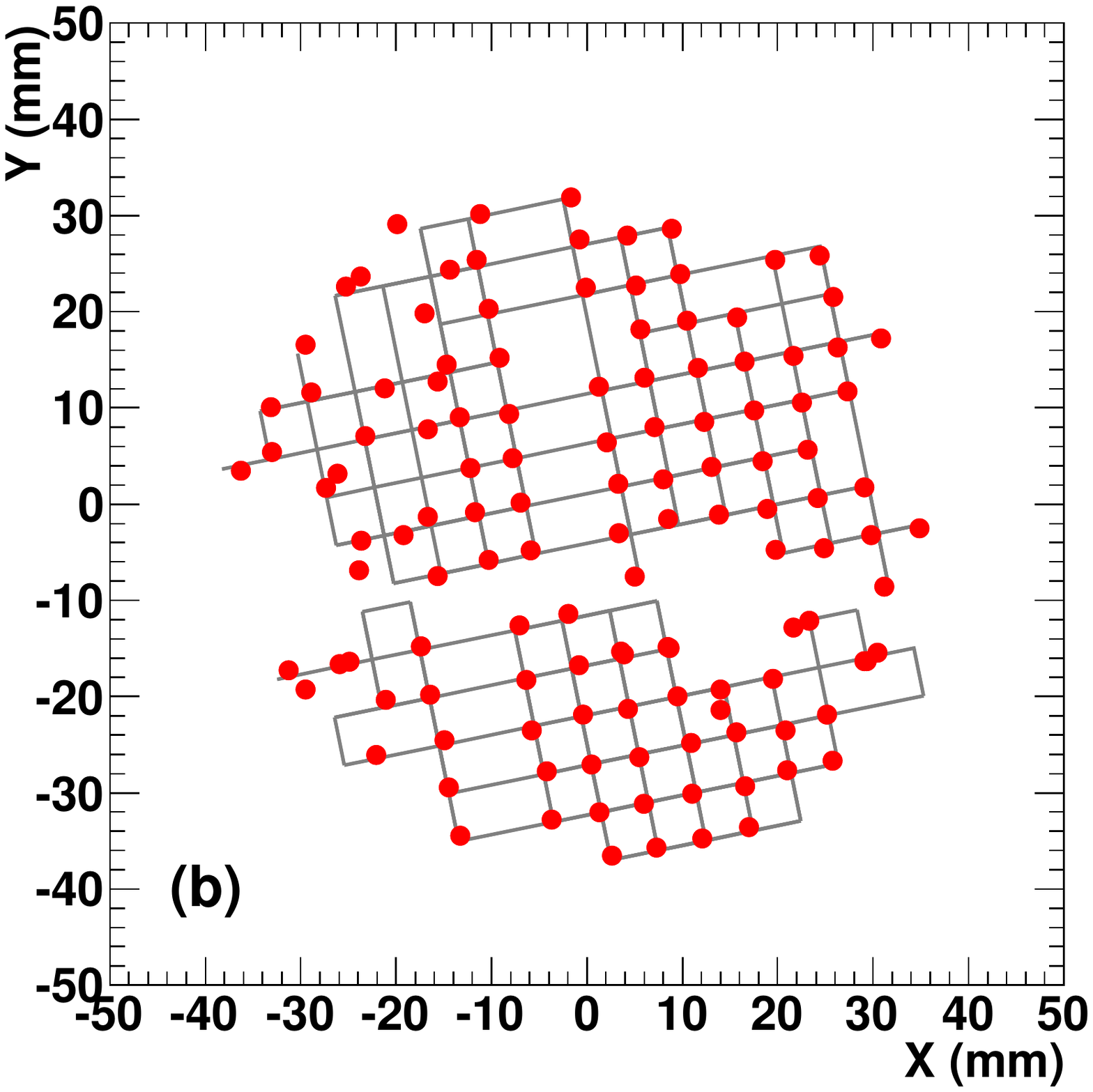}
 \label{fig:parfit_corr_2grids}}
\caption{MDCS distortion correction applied to a typical HPD: (a) before distortion correction, and (b) after distortion correction.  The cleaned peak positions at the photocathode before and after correction are shown as dots. The expected grid patterns from two LED units are superimposed.}
\label{fig:parfit_2grids}
\end{center}
\end{figure}

\subsection{Pattern Cleaning}

A pattern-cleaning algorithm is used to distinguish primary signal peaks from background arising from the above sources.
Four variables are used to characterize the properties of the peaks and to build a likelihood that optimally selects signal.
These variables are:
(1) the height of the fitted 2D peak,
(2) the number of hits recorded across the HPD in which the peak is found,
(3) the proportion of these hits that contribute to the peak,
and
(4) the displacement of the peak in the direction of the light-bar's travel, compared to the average displacement of all the peaks found in the other HPDs in the column.

An unbinned maximum likelihood fit is performed on these four variable distributions and combined into a likelihood for the signal and the background hypotheses.
The logarithmic difference between the two likelihoods (Delta Log Likelihood: DLL) is taken as the discriminant variable for the peak selection.
The DLL distributions for single and non-single peaks is shown in Fig.{~\ref{fig:MDMSDLLdistrib}}.
As expected, HPDs which have a single fitted peak have a high likelihood of being a primary signal peak.
The cut in DLL can be varied depending on the purity required, however usually a $DLL > 0$ cut is applied.
If more than one peak passes the DLL cut, the one with the highest DLL is kept.
With this technique, a large fraction of useable peaks are retained from HPD images containing multiple peaks by requiring $DLL > 0$ (HPD images with a single peak are unambiguous and always retained). An example of the cleaned pattern is shown in Fig.{~\ref{fig:final_peaks}}. The resulting set of cleaned peaks is used to extract the final calibration parameters.

\section{Calibration Extraction}

The distortion of the photoelectron image in the presence of the magnetic field is observed to be modest and smoothly varying.
This allows an inverse mapping strategy, mentioned above, to be implemented via a parameterization of the distortion rather than a look-up table procedure.
Parameters are obtained for each HPD and stored in an LHCb database for use in the Cherenkov ring reconstruction with collision data.

\begin{figure}[thb]
 \begin{center}
 \includegraphics[width=6cm,bb=100 0 450 450]{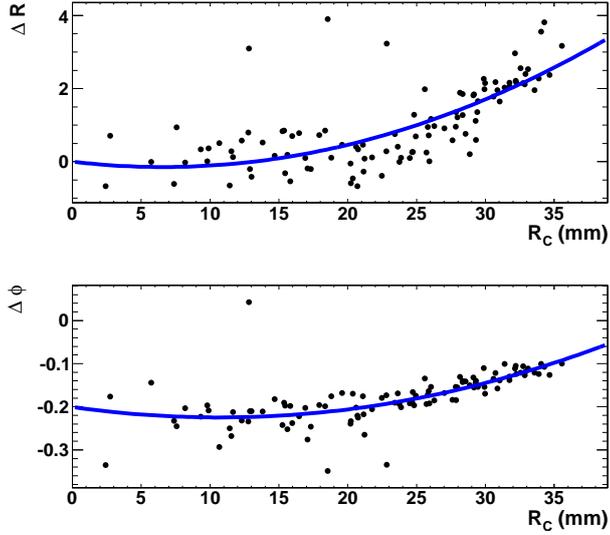}
 \caption{Polynomial fits of distortion pairs $\Delta R$ and $\Delta \phi$ as a function of $R_C$.}
 \label{fig:deltafitted_example}
\end{center}
\end{figure}

\subsection{Parameterization}

Due to design of the RICH and its shielding, the magnetic field inside an the HPDs is, to a good approximation, parallel to the tube axis.
The absence of a significant transverse field simplifies the parameterization, as radial symmetry about the tube axis can be assumed.
The transformation of the position of a peak on the silicon chip to that of the corresponding light spot at the exterior of the HPD window involves both radial and axial components.
Any point on the pixel chip has coordinates $(R_A,\phi_A)$, viz., the radius and azimuthal angle with respect to the center of radial symmetry on the chip (chip center).
Similarly, the corresponding point on the outer surface of the quartz window has coordinates $(R_C,\phi_C)$.
The transformation is parameterized as
\begin{equation}
\begin{array}{rcl}
R_C &=& \rho_1 R_A + \rho_2 R^2_A + \rho_3 R^3_A \\
\phi_C &=&  \phi_A + \theta_0 + \theta_1 R_A + \theta_2 R^2_A + \theta_3 R^3_A  \,\,  ,
\label{eqn:trasf_condb}
\end{array}
\end{equation}
where $\rho_i$ and $\theta_i$ are the parameters to be extracted.
The radial part is a combination of the electrostatic magnification, the optical refraction at the curved surfaces of the photocathode window, and the magnetic distortion correction.
The axial part describes the characteristic ``S-shaped'' displacement, as is evident in Fig.~\ref{fig:cleaned_peaks}.
These parameters are expected to depend only on the magnetic field strength and to be stable in time.
The whole photocathode image (and so, the chip center) can move if small amounts of charge build up on insulating surfaces within the tube.
This is measured periodically by fitting a circle to the uniformly-illuminated HPD disk image from proton-proton collision data.
This procedure automatically takes into account misalignment of the chip with respect to the tube axis.

\subsection{Fitting Procedure}

The calibration parameters are deduced by minimizing the distance between the position of the selected peaks, suitably transformed with Eq.~\ref{eqn:trasf_condb}, and the cartesian grid that is defined by the geometry of the light bar and the logic of its movement.
Once cleaned, only one peak per HPD remains; so it is trivial to associate it with the originating light spot, and hence provide strong constraints in the fit.
As the LED unit is shorter than the diameter of the HPD tube, an HPD is typically illuminated by two LED PCB units which result in two distinguishable grid patterns.
Depending on the step number within the LED illumination sequence, the light may come from one unit or the other, but not from both in the same step.
Therefore these two distinct grids are easily separated in the data, as seen in Fig.~\ref{fig:parfit_2grids}.
The spacing between light spots at the photocathode and the spacing between two grids (inter-LED unit distance) is imposed using the design specifications of the light bar.
The absolute positions of the light spot grids in the $xy$ plane is not imposed but fitted together with the distortion parameters to allow for a global misalignment of the HPD tube.

Each peak position and corresponding grid point form a ``pair''.
The pair is described by the parameters $\Delta R_C$ and $\Delta \phi_C$: orthogonal distances between the two points when projected onto a 2D surface perpendicular to the cylindrical axis of the HPD.
Pairs with $\Delta R_C$ and/or $\Delta\phi_C R_C$ greater than twice the spacing between two consecutive LEDs are not considered in the fit, providing additional protection from residual backgrounds.
An example of the relation of $\Delta R_C$ and $\Delta\phi_C$ versus radius $R_C$ is shown in Fig.{~\ref{fig:deltafitted_example}}.
These shapes are parameterized by simple polynomial functions \cite{AR2004,AR2005}, and then combined with a coefficient of magnification to arrive at the full transformation, Eq.~\ref{eqn:trasf_condb}.
The peak positions after the full transformation are shown in Fig.{~\ref{fig:parfit_corr_2grids}} and exemplify the validity of the parameterization.

\section{Performance on Calibration Data}

\subsection{Position Resolution}

\begin{figure}[bht]
 \begin{center}
 \subfigure{\includegraphics[width=5cm,bb=100 0 450 384]{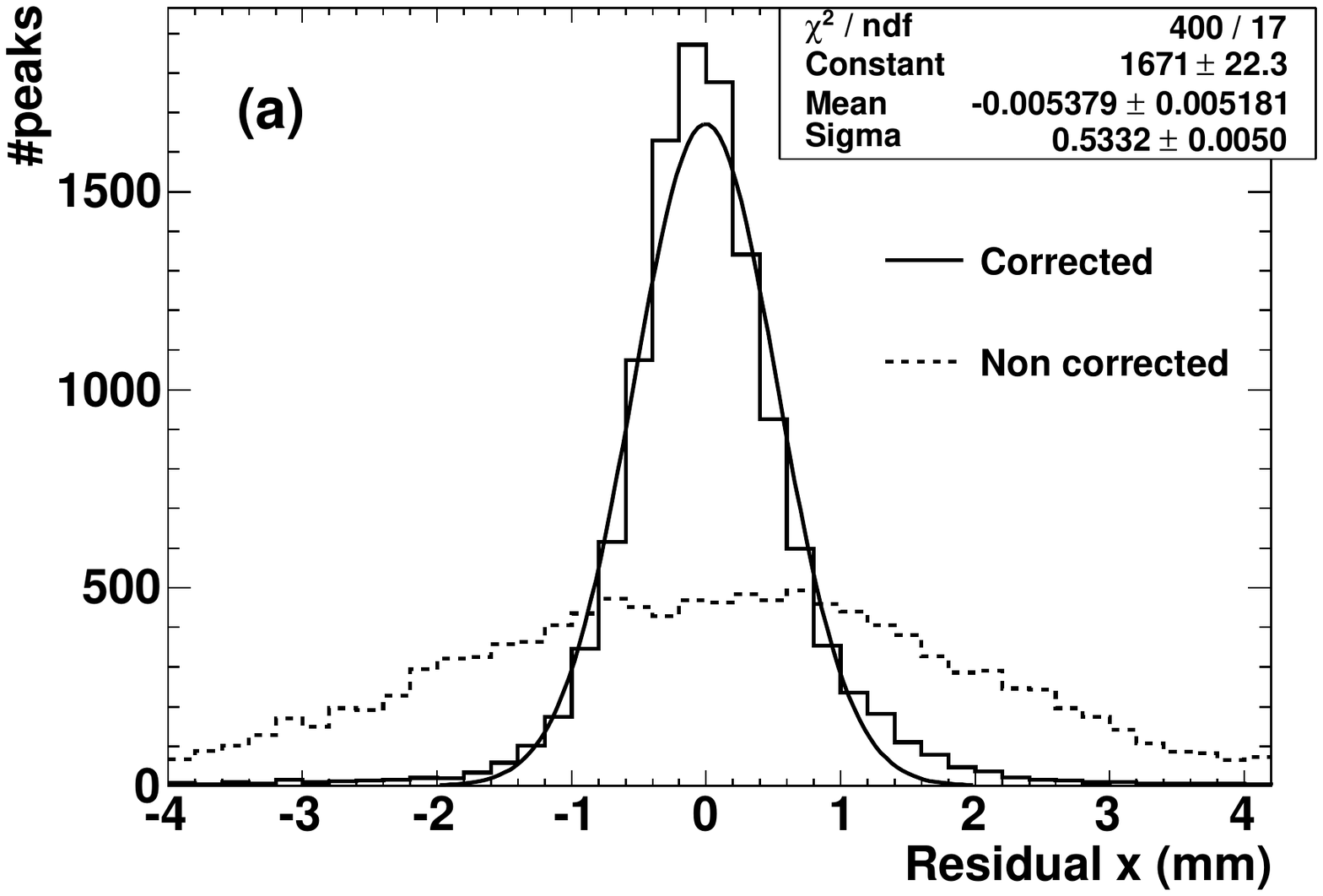}}
 \subfigure{\includegraphics[width=5cm,bb=100 0 450 384]{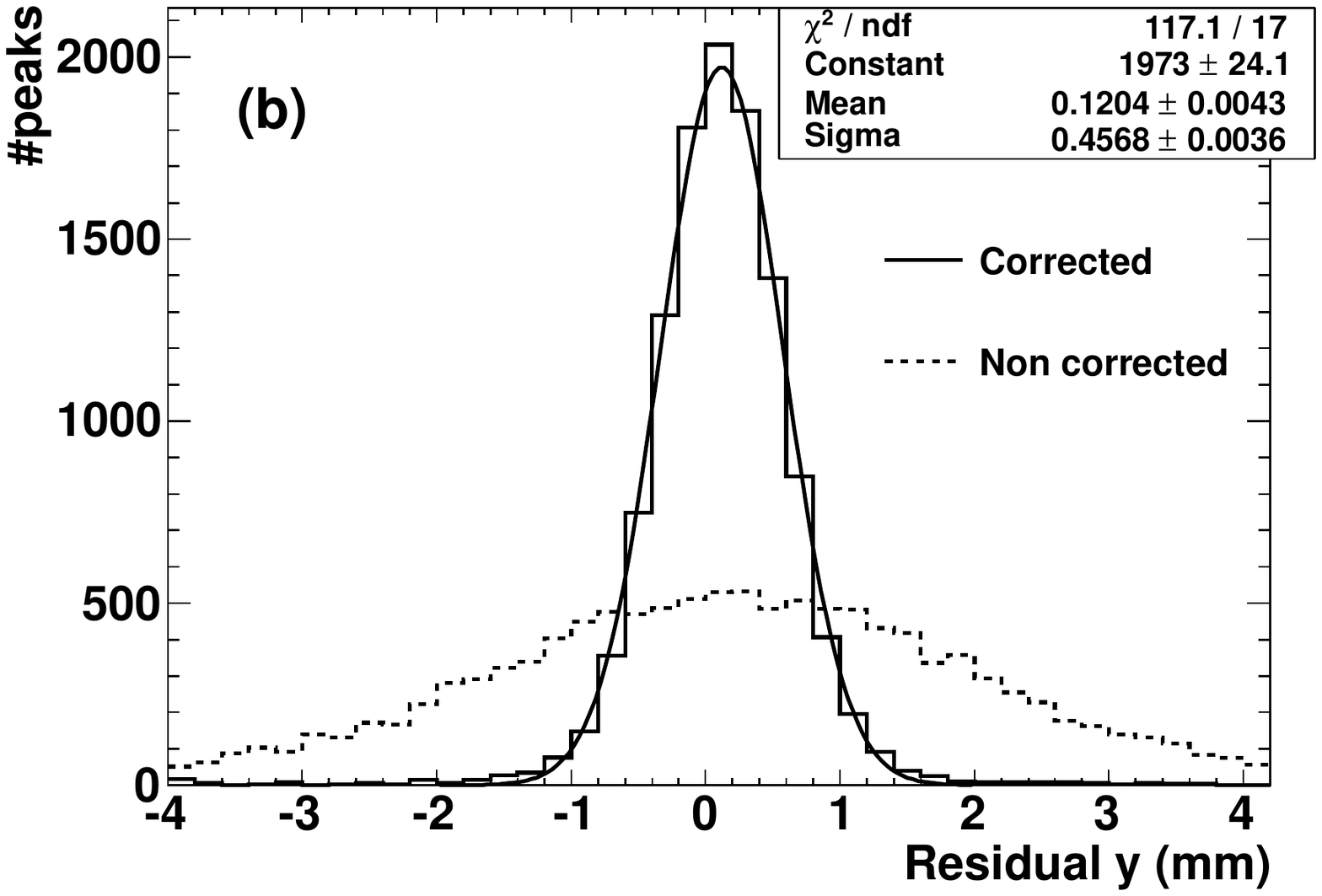}}
 \caption{Typical position resolution of corrected peak positions in calibration data, as measured at the HPD quartz window, for (a) x (LHCb pixel) direction, and (b) y (ALICE pixel) direction.  The uncorrected peak positions are indicated by the dashed histogram.}
 \label{fig:resol_mdcs}
\end{center}
\end{figure}

The distribution of residuals between the distortion-corrected peaks and the expected LED position is shown in Fig.{~\ref{fig:resol_mdcs}}.
Clear benefit is seen from application of the fitted parameterization, particularly in the ALICE pixel direction, which has the higher granularity.
The variance of the post-transformation data can be interpreted as the residual uncertainty on the position of incident photons due to the magnetic distortions.
A simple Gaussian fit gives a width $\sigma\approx0.49$~mm (outside the HPD quartz window) which corresponds to an factor of $\sim$3.7 improvement with respect to the uncorrected distribution.
This remaining uncertainty compares well with that due to the binary resolution of the LHCb pixels ($0.81$~mm).
So it can be concluded that the residual magnetic distortions inflate the effective pixel contribution to the Cherenkov angle resolution by just $\sim$17\%.

\begin{figure}[thb]
 \begin{center}
 \subfigure{\includegraphics[width=5cm,bb=100 0 450 380]{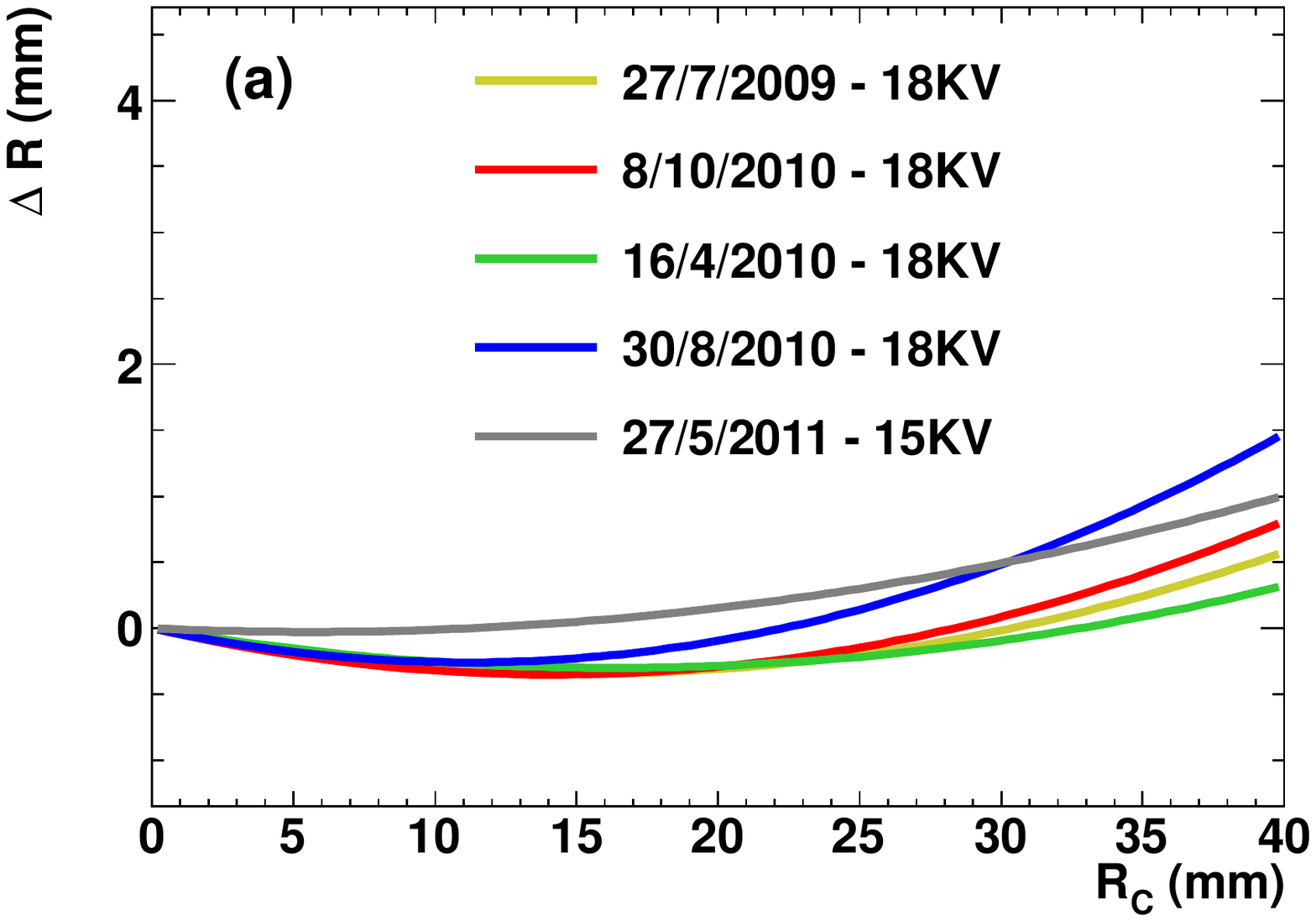}}
 \subfigure{\includegraphics[width=5cm,bb=100 0 450 380]{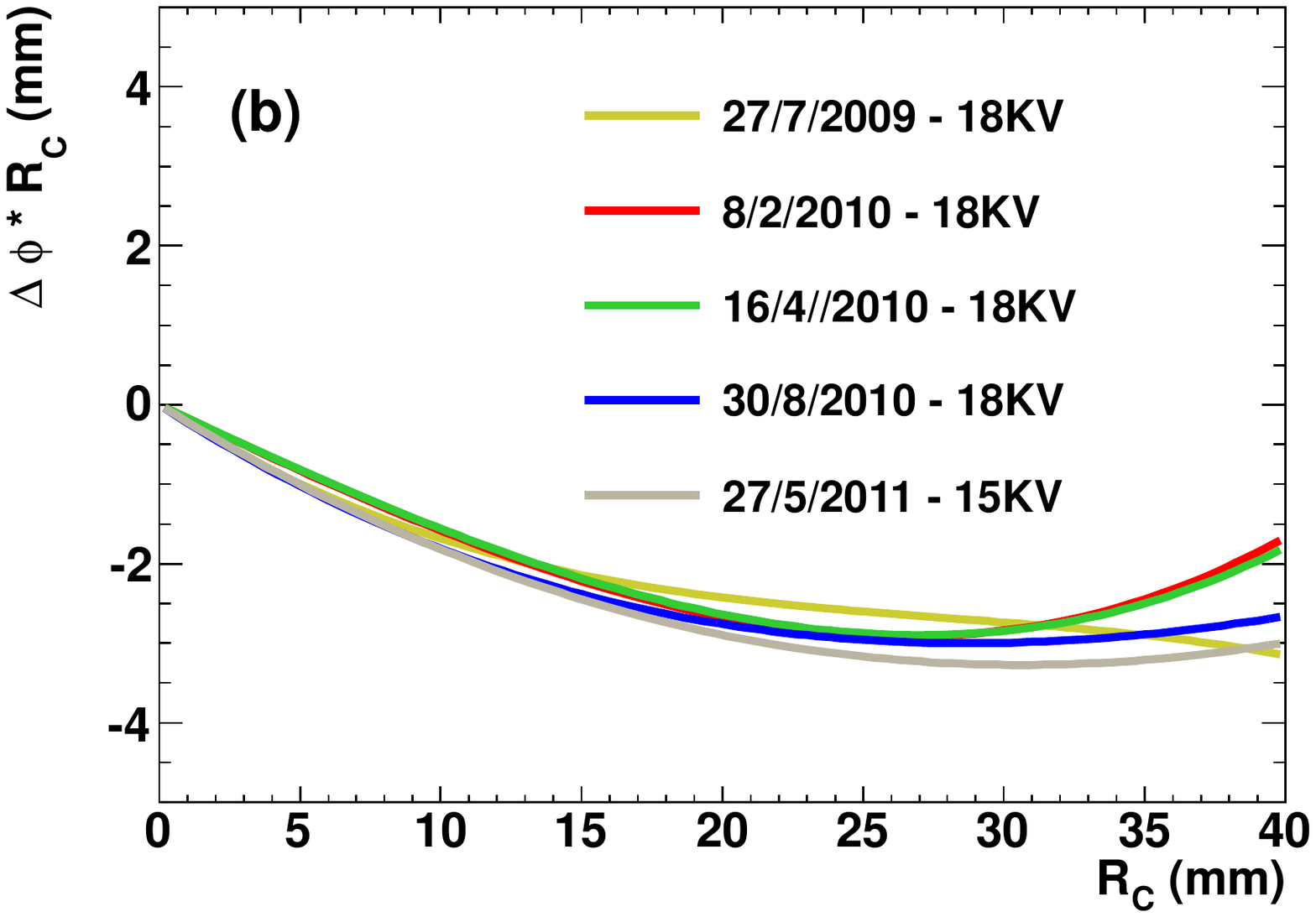}}
 \caption{Magnetic distortion correction to the magnification for different scans and HPD high voltages over a two year period, for a typical HPD. (a) Radial correction. (b) Axial correction.}
 \label{fig:trends_func}
\end{center}
\end{figure}

\begin{figure}[t]
\begin{center}
 \subfigure{\includegraphics[width=6cm,bb=100 0 450 512]{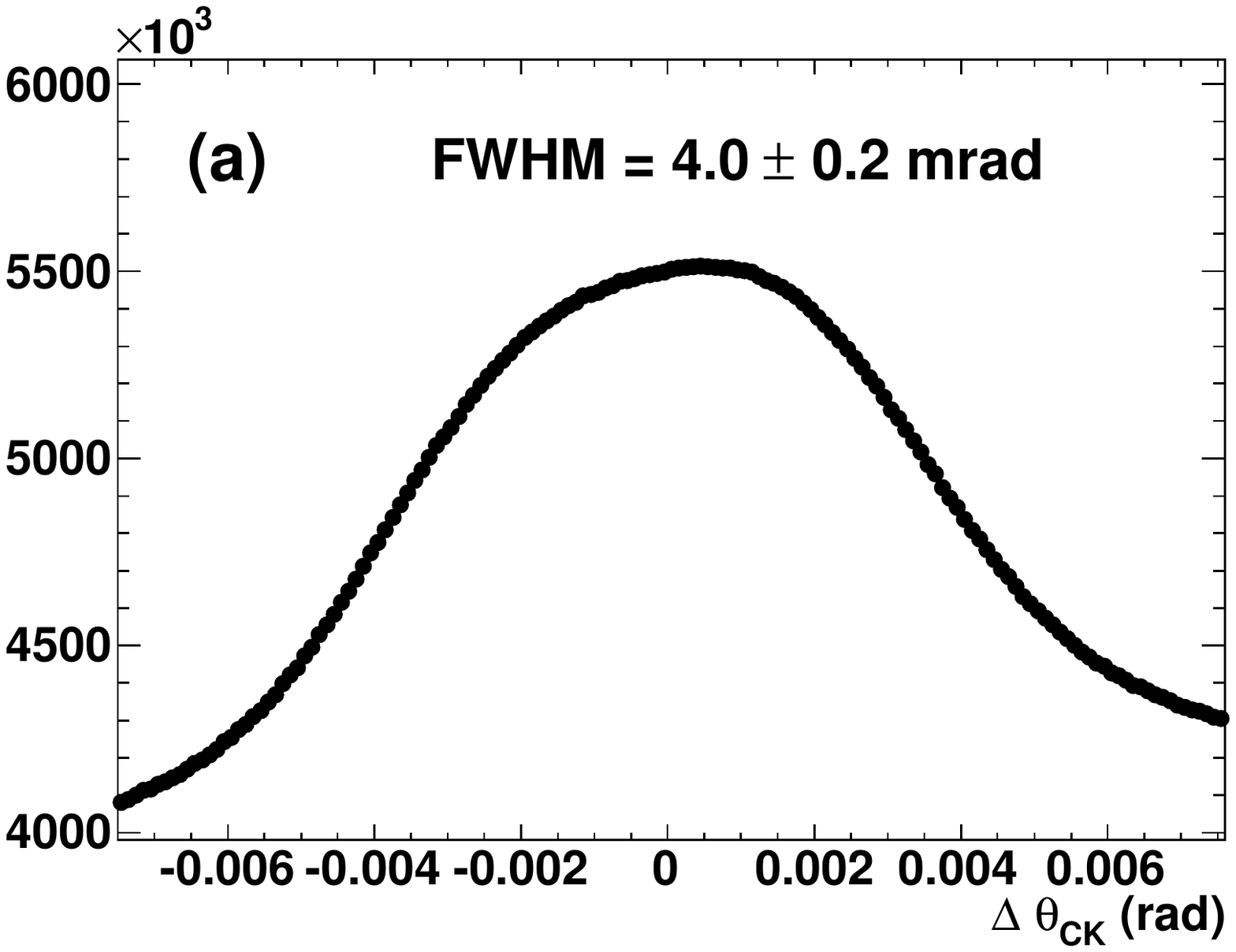}
 \label{fig:CK_resol11_no_mdcs}}
 \subfigure{\includegraphics[width=6cm,bb=100 0 450 384]{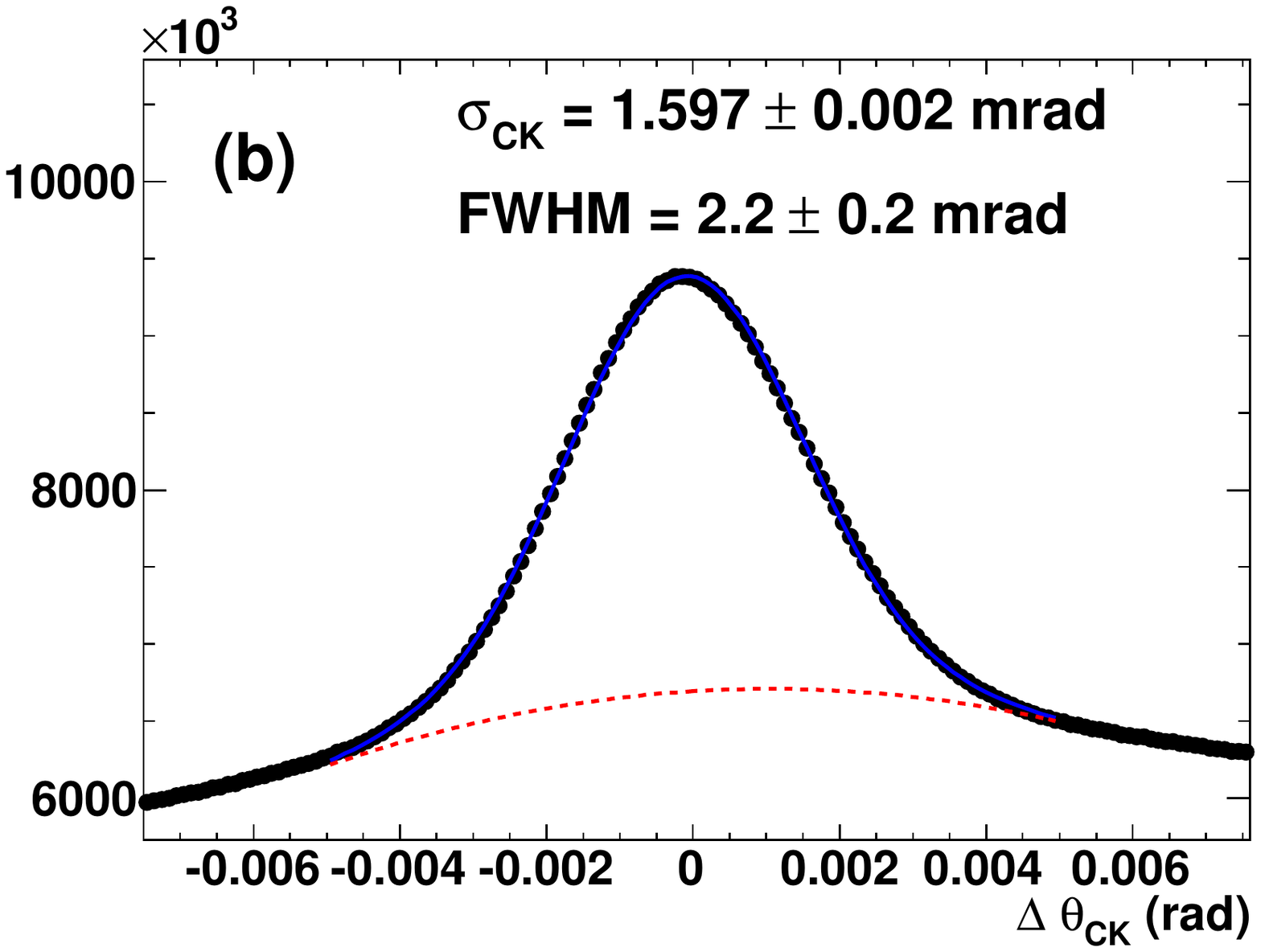}
 \label{fig:CK_resol11}
}
 \caption{The per-photon Cherenkov angle resolution in 2011 collision data: (a) before MDCS distortion correction applied, and (b) after MDCS distortion correction applied.}
 \label{fig:CK_res_all}
\end{center}
\end{figure}

\subsection{Magnification}

The demagnification power of the HPDs is extracted from these data and compared with the electrostatic design expectation of a factor five.
In the absence of magnetic distortions (using data taken whilst the spectrometer magnet is off) the effective demagnification of the HPDs, including the fish-eye refraction of the curved photocathode window, is measured.

A second order polynomial is sufficient to parameterize the magnification plus refraction effects.
The total magnification factor, averaged over all HPDs, is 5.6.
The linear term dominates, having an average value 5.3, with the quadratic term becoming significant at high $R_C$, near the perimeter of the quartz window.

\subsection{Correction Stability}

The stability of the MDCS corrections over time is related to HPD behavior and long-term MDCS performance.
Fig.{~\ref{fig:trends_func}} shows the axial and radial magnetic distortion corrections (with magnification constant) for a particular HPD at different times and high voltage conditions over nearly a two year period. For the entire $R_C$ range, the radial and axial corrections are stable and within the resolution of the method.
This legitimates the use of the same calibration parameters during several months of experimental operation.
A new calibration is performed in the event of HPD replacement or significant changes in operation (e.g., change of high voltage).

\section{Performance on Collision Data}

The most important performance metric of the MDCS system
is its effect on the RICH1 Cherenkov angle resolution in proton-proton collision data.
For saturated ($\beta=1$) Cherenkov rings in collision events with hadrons in the final state,
Fig.{~\ref{fig:CK_res_all}} shows the per-photon Cherenkov angle resolution for the same data, with and without the MDCS corrections applied.
With the MDCS corrections, a fit with a single Gaussian plus a second order polynomial background gives a resolution of $1.597\pm0.002$ mrad.  This is close to the expected value of 1.55 mrad from simulation.
Without the MDCS corrections, the resolution is about a factor of two worse, as determined by a comparison of the FWHM.

\section{Conclusions}

The LHCb RICH1 magnetic distortion calibration system described here is dedicated to correcting the photon positions from the effects of magnetic fields and other distortions in the HPD photons detectors.  The MDCS system has been in operation since the start of LHCb data-taking. An analysis method has been developed to extract distortion correction coefficients from calibration data.
Applying these corrections to proton-proton collision data, the per-photon Cherenkov resolution angle resolution is improved by a factor of two, and now approaches the expectation from simulation.
This improvement greatly enhances the particle identification performance of the LHCb experiment, and
indicates that the MDCS system has fulfilled its design requirements.

\section{Acknowledgements}

The authors would like to thank
L.~Buda, P.~Arnold, L.~Schmutzler (Syracuse),
R.~Plackett, T.~Savidge (Imperial),
K.~Wyllie, D.~Piedigrossi, T.~Gys, A.~Cherif, W.~Witzeling (CERN),
and
A.~Papanestis (U.K.\ Science and Technology Facilities Council),
for their help in this project.
In particular, the authors thank
C.~Brown (Syracuse) for his herculean effort in the machining and construction phases,
and P.~Gelling (SenSyr LLC) for superior electronic design and fabrication work.

We thank the U.S.\ National Science Foundation
and the U.K.\ Science and Technology Facilities Council
for support.






\bibliographystyle{model1-num-names} 
\bibliography{MDCS}







\end{document}